\documentclass[a4paper,fleqn,usenatbib]{mnras}
\usepackage{float}
\usepackage[T1]{fontenc}
\usepackage{ae,aecompl}
\usepackage{graphicx}
\usepackage{amsmath}
\usepackage{amssymb}
\usepackage{supertabular}
\usepackage{siunitx}
\usepackage{caption}
\usepackage{longtable}
\usepackage{geometry}
\def\beq{\begin{eqnarray}}
\def\eeq{\end{eqnarray}}

\title[BH growths in GRBs driven by BZ mechanism]{Black hole growths in gamma-ray bursts driven by the Blandford-Znajek mechanism}
\author[Li \& Liu]{Xiao-Yan Li, and Tong Liu\thanks{E-mail: tongliu@xmu.edu.cn}\\
Department of Astronomy, Xiamen University, Xiamen, Fujian 361005, China}
\date{Accepted XXX. Received YYY; in original form ZZZ}
\pubyear{2023}

\begin{document}

\label{firstpage}
\pagerange{\pageref{firstpage}--\pageref{lastpage}}
\maketitle

\begin{abstract}
The Blandford-Znajek (BZ) mechanism in stellar-mass black hole (BH) hyperaccretion systems is generally considered to power gamma-ray bursts (GRBs). Based on observational GRB data, we use the BZ mechanism driven by the BH hyperaccretion disc to investigate the evolution of the BH mass and spin after the jets break out from the progenitors. We find that the BH growths are almost independent of initial BH masses. Meanwhile, the BH growths will be more efficient with smaller initial spin parameters. We conclude that (i) the BZ mechanism is efficient for triggering BH growths for only 1 of 206 typical long-duration GRBs; (ii) the mean BH mass growths of ultra-long GRBs are marginal for all 7 samples collected; (iii) for the short-duration GRBs, the results that BHs show minimal growths is consistent with the mass supply limitation in the scenario of compact object mergers.
\end{abstract}

\begin{keywords}
accretion, accretion discs - black hole physics - gamma-ray burst: general - magnetic fields
\end{keywords}

\section{Introduction} \label{sec:introduction}

Gamma-ray bursts (GRBs) are the most luminous transients associated with gravitational waves, neutrinos, and cosmic rays known in the universe. Based on the GRB prompt duration $T_{90}$, defined as the interval when $5\% \mbox{-}95\%$ of the burst fluence is accumulated, GRBs are classified as long-duration GRBs (LGRBs; $T_{90}>2\,\mathrm{s}$) and short-duration GRBs (SGRBs; $T_{90}<2\, \mathrm{s}$) \citep[e.g.,][]{1993ApJ...413L.101K} or Type I and II GRBs \citep[e.g.,][]{2006Natur.444.1010Z,2007ApJ...655L..25Z}. Some LGRBs are confirmed to be associated with Type Ib/c supernovae \citep[SNe, see e.g.,][]{2003Natur.423..847H,2003ApJ...591L..17S,2004ApJ...609L...5M,2011ApJ...743..204B}. The association of some LGRBs with SNe suggests that LGRBs originated from the collapses of massive stars \citep[$>8\, M_{\odot}$, e.g.,][]{1993ApJ...405..273W,1998ApJ...494L..45P}. SGRBs are generally believed relevant to the mergers of compact objects, i.e., a binary neutron star (NS) or black hole (BH)-NS \citep[e.g.,][]{1989Natur.340..126E,1992ApJ...395L..83N,2007PhR...442..166N}. Interestingly, ultra-long GRBs (ULGRBs; $T_{90}>1000\mathrm{~s}$) have been observed in the past decade \citep[e.g.,][]{2013ApJ...766...30G,2014ApJ...781...13L}. Some literatures proposed that ULGRBs may belong to the third population and have novel progenitors \citep[e.g.,][]{2014ApJ...781...13L,2015Natur.523..189G}. On the other hand, the lack of compelling evidences prevents a new classification in the burst-duration distribution \citep[e.g.,][]{2013ApJ...778...54V,2014ApJ...787...66Z}. In addition, ULGRBs are associated with SNe, which implies that they might be produced by the massive collapsars \citep[e.g.,][]{2015Natur.523..189G,2018ApJ...852...20L}.

For both collapse and merger scenarios, a rotating BH surrounded by a hyperaccretion disc \citep[e.g.,][]{1991AcA....41..257P,1999ApJ...524..262M,2017NewAR..79....1L} or a megnetar \citep[e.g.,][]{1992ApJ...392L...9D,1992Natur.357..472U,1998A&A...333L..87D,1998PhRvL..81.4301D,2011MNRAS.413.2031M} can be formed. In the hyperaccretion system, the violent accretion powers an energetic engine, then the neutrino annihilation processes \citep[e.g.,][]{1997A&A...319..122R,1999ApJ...518..356P,2003MNRAS.345.1077R,2007ApJ...661.1025L,2011MNRAS.410.2302Z,2013ApJ...766...31K} or the Blandford-Znajek (BZ) mechanism \citep[e.g.,][]{1977MNRAS.179..433B,2000PhR...325...83L,2015ApJS..218...12L} can produce the ultrarelativistic jets to trigger GRBs \citep[for a review, see][]{2017NewAR..79....1L}. A neutrino-dominated accretion flow (NDAF) around a stellar-mass BH with an extremely high accretion rate is a plausible candidate for the central engine. In this scenario, neutrinos exploit the thermal energy of the heated disc and liberate enormous binding energy, and then annihilate outside of the disc to produce the fireball. The BZ mechanism is an alternative model to drive GRB jets, in which case the BH's rotational energy are efficiently extracted by the Poynting jets.

Moreover, the evolution of BH mass and spin may be violent in the hyperaccretion system \citep[e.g.,][]{2015ApJ...806...58L,2015ApJ...815...54S,2022ApJ...929...83Q}. If the initial BH mass is set as to approximately $3\,M_\odot$, according to the neutrino annihilation process, the final BH mass is relatively large ($>5\,M_\odot$) for LGRBs and small ($3-5 \,M_\odot$) for SGRBs. The BZ mechanism is less efficient in triggering BH growths compared with the neutrino annihilation processes \citep[e.g.,][]{2015ApJS..218...12L,2022ApJ...929...83Q}. However, in the scenarios of massive collapsars and compact object mergers, the neutrino annihilation process could only last for several or tens of seconds until the ignition accretion cannot be reached with the decreasing accretion rate. Then it might be quickly replaced by the BZ mechanism and jets could break out from the envelopes or ejecta, especially for collapsar model \citep[e.g.,][]{2018ApJ...852...20L,2021MNRAS.507..431W}. Moreover, the typical luminosity of the neutrino annihilation process is lower than that of the BZ mechanism by about two orders of magnitude with the same high spin parameter and accretion rate \citep[e.g.,][]{2015ApJS..218...12L}. If the disc outflows are considered, the above conclusion will be strengthened. Of course, one cannot rule out that the neutrino annihilation process will be dominated in the activities of GRB central engines \citep[e.g.,][]{2017NewAR..79....1L}. Here we generally discussed the BH growths of GRBs powered by the BZ mechanism and set the initial BH properties at the moment that the jets break out form the envelopes of the massive collapsars or the ejecta of the compact object mergers. The growth caused by the BZ mechanism is small but not totally negligible, which is directly related to the observable GRBs as mentioned above. The growths of the BHs in the center of GRBs is an inevitable course in the evolution history of stellar-mass BHs, then effects the mass distribution of compact objects \citep[e.g.,][]{2021ApJ...908..106L}.

In this paper, combining with the GRB samples, we visit the BH hyperaccretion system with the BZ mechanism to test the suitability of the BZ mechanisam for GRBs and investigate the growths of newborn BHs in the center of GRBs. This paper is organised as follows. In Section 2, the model for the evolution of BH characteristic parameters is described. The results are shown in Section 3. Conclusions and discussion are presented in Section 4.

\section{Model} \label{sec:model}

\begin{figure}
\centering
\includegraphics[angle=0,width=0.5\textwidth]{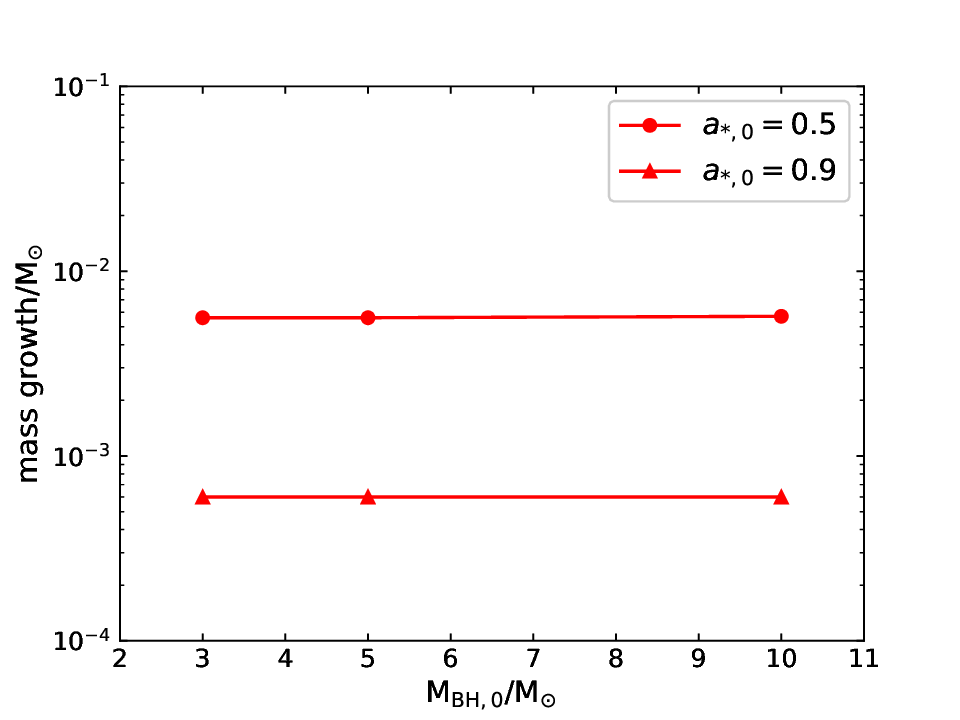}
\caption{Influence of initial BH masses and spin parameters on BH mass growths in a typical GRB case with the luminosity $L_{\mathrm{j}}=10^{49} ~\mathrm{erg}\,\mathrm{s}^{-1}$ and duration $T_{90}=30\,\mathrm{s}$. The circles and triangles represent the initial BH spin $a_{*,0}= 0.5$ and $0.9$, respectively.}
\label{MyFigA}
\end{figure}

The characteristic parameters of the BH will undergo drastic evolution if a rotating stellar BH surrounded by a hyperaccretion disc with a very high accretion rate is the central engine of GRBs \citep[e.g.,][]{2015ApJ...806...58L,2015ApJ...815...54S,2022ApJ...929...83Q}. Based on the conservation of the mass and angular momentum, considering that part of the BH rotational energy would be extracted by the Poynting jet for the BZ mechanism, the mass and angular momentum of a Kerr BH evolved with time read as \citep[e.g.,][]{2000PhR...325...83L,2000JKPS...36..188L}
\begin{equation}
    \frac{d M_{\mathrm{BH}}}{d t}=\dot{M} e_{\mathrm{ms}}-\frac{L_{\mathrm{BZ}}}{c^2},
\label{eq1}
\end{equation}
\begin{equation}
    \frac{d J_{\mathrm{BH}}}{d t}=\dot{M} l_{\mathrm{ms}}-\frac{L_{\mathrm{BZ}}}{c^2 \Omega_{\mathrm{F}}},
\label{eq2}
\end{equation}
where $M_{\rm BH}$ and $J_{\rm BH}$ are the mass and angular momentum of the BH, $\dot{M}$ is the mass accretion rate, $c$ is the speed of light, and $e_{\mathrm{ms}}$ and $l_{\mathrm{ms}}$ are the specific energy and angular momentum corresponding to the marginally stable orbit radius of the BH. They are defined as $e_{\mathrm{ms}}=\frac{1}{\sqrt{3x_{\mathrm{ms}}}}\left(4-\frac{3 a_{*}}{\sqrt{x_{\mathrm{ms}}}}\right)$ and $l_{\mathrm{ms}}=2 \sqrt{3} \frac{G M_{\mathrm{BH}}}{c}\left(1-\frac{2 a_{*}}{3 \sqrt{x_{\mathrm{ms}}}}\right)$, where $a_{*} \equiv c J_{\mathrm{BH}} / G M_{\mathrm{BH}}^{2}$ ($0\leq a_{*}\leq 1$) is the dimensionless spin parameter of the BH and $x_{\rm{ms}}$ is the dimensionless marginally stable orbit radius of the BH, which can be written as $x_{\mathrm{ms}}=3+Z_{2}-\sqrt{(3-Z_{1})(3+Z_{1}+2 Z_{2})}$ with $Z_{1}=1+(1-a_{*}^{2})^{1/3}[(1+a_{*})^{1/3}+(1-a_{*})^{1/3}]$ and $Z_{2}=\sqrt{3 a_{*}^{2}+Z_{1}^{2}}$ \citep[e.g.,][]{1972ApJ...178..347B,1998GrCo....4S.135N,2008bhad.book.....K}. The last two terms on the right-hand side of Equations (1) and (2) are relevant to the BZ mechanism, where $L_{\mathrm{BZ}}$ is the BZ jet power and $\Omega_{\mathrm{F}}$ is the magnetic field angular velocity at the marginally stable orbit radius. Here, we adopt the optimal mode $\Omega_{\mathrm{F}}=\Omega_{\mathrm{H}}/2$ \citep[e.g.,][]{2000PhR...325...83L,2000JKPS...36..188L}, with $\Omega_{\mathrm{H}} \equiv a_{*} c^3 / [2 (1+\sqrt{1-a_{*}^2}) GM_{\mathrm{BH}}]$ being the angular velocity on the stretched horizon.

\begin{figure}
\centering
\includegraphics[angle=0,width=0.49\textwidth]{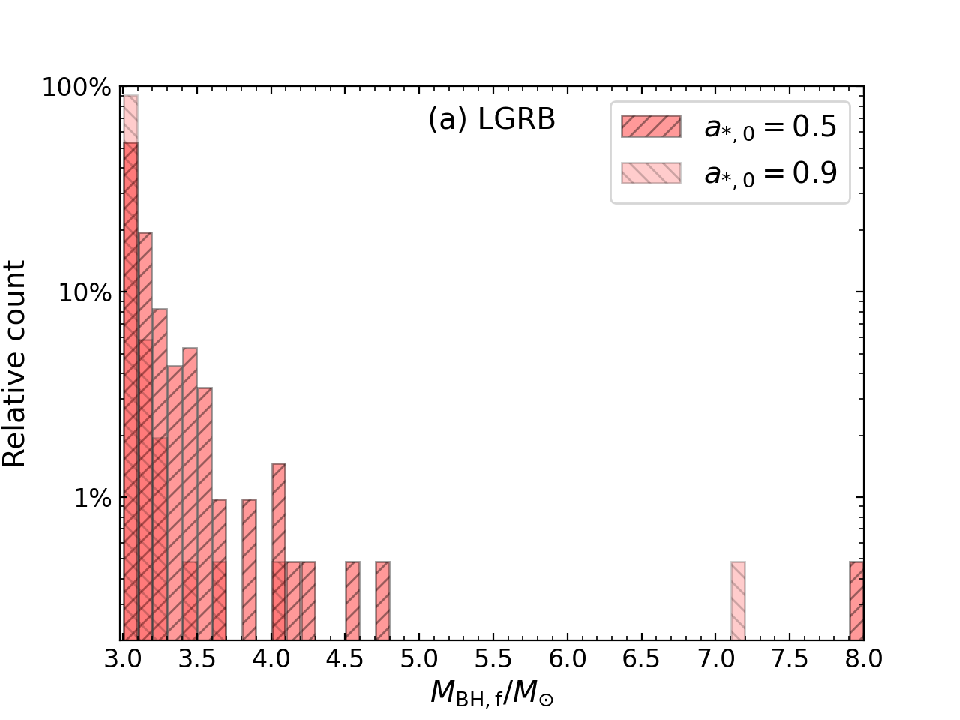}
\includegraphics[angle=0,width=0.49\textwidth]{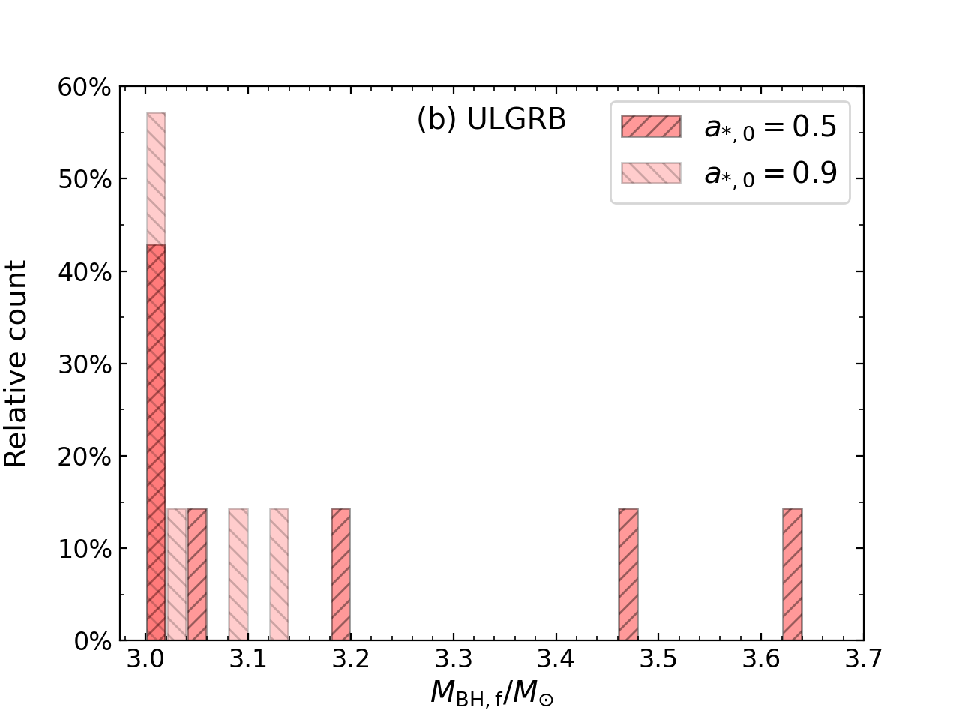}
\includegraphics[angle=0,width=0.49\textwidth]{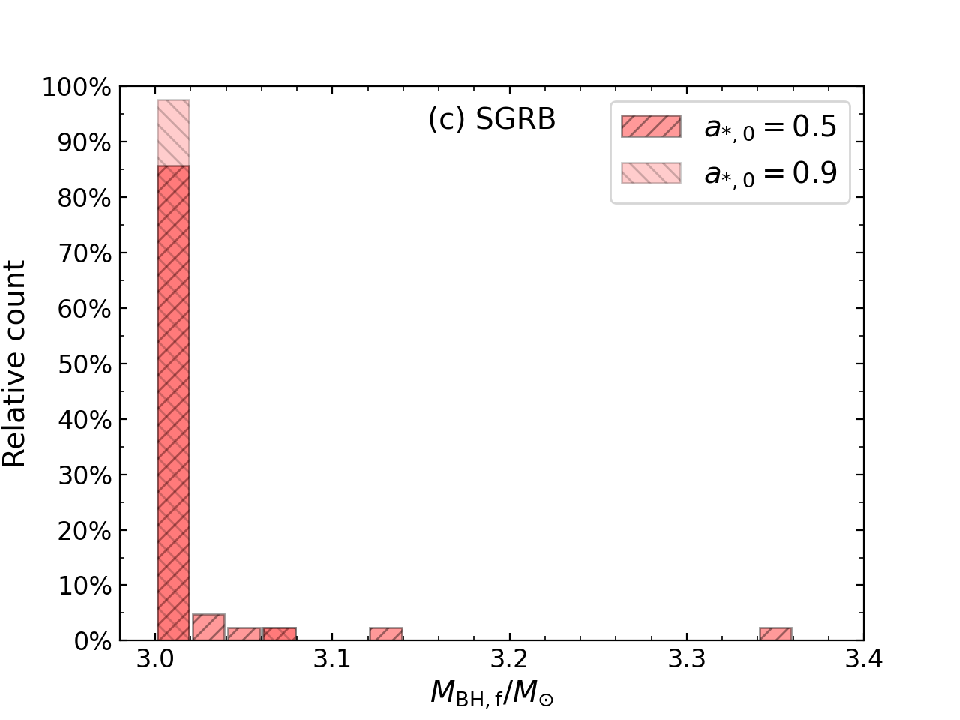}
\caption{Distributions of the final BH masses $M_{\rm BH,f}$. The initial BH mass $M_{\rm BH,0}$ is 3 $M_\odot$. The dark and light colors denote $a_{*,0}= 0.5$  and $0.9$, respectively.}
\label{MyFigB}
\end{figure}

\begin{figure}
\centering
\includegraphics[angle=0,width=0.49\textwidth]{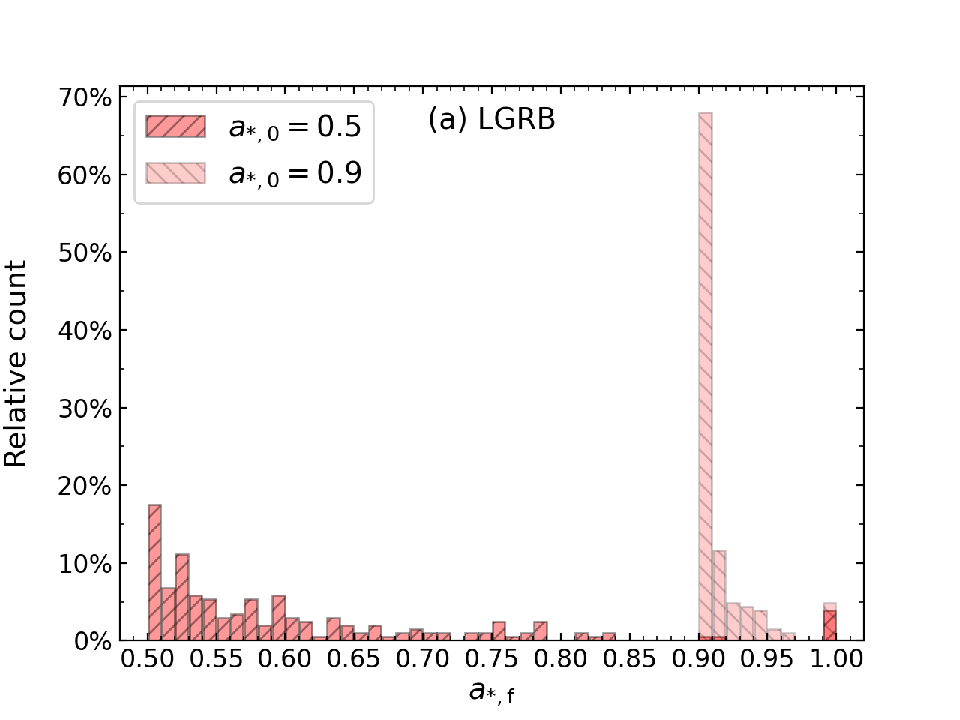}
\includegraphics[angle=0,width=0.49\textwidth]{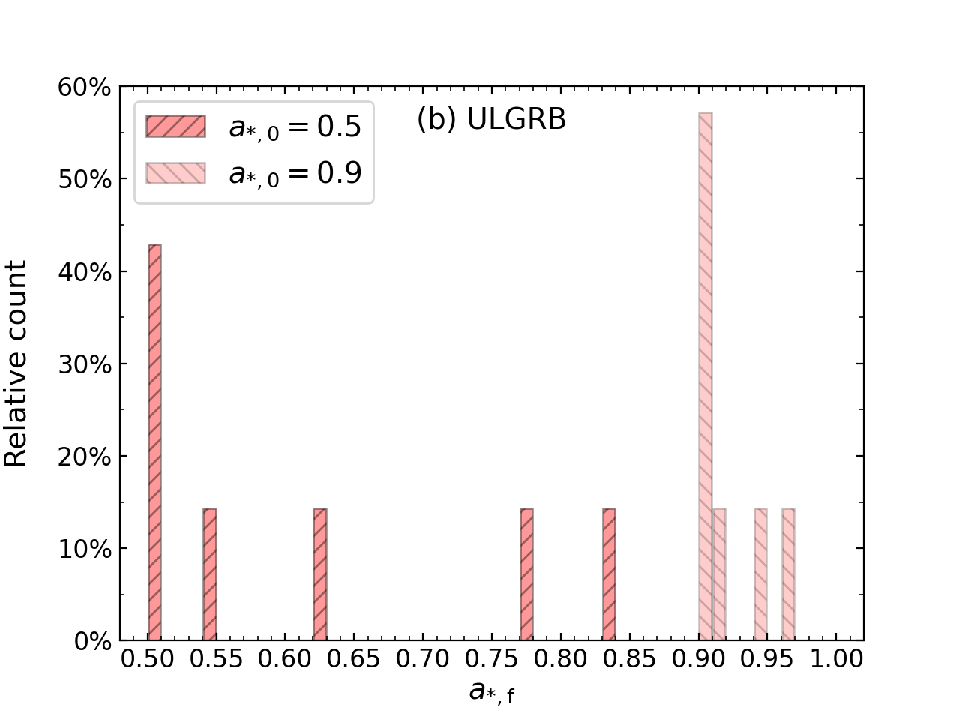}
\includegraphics[angle=0,width=0.49\textwidth]{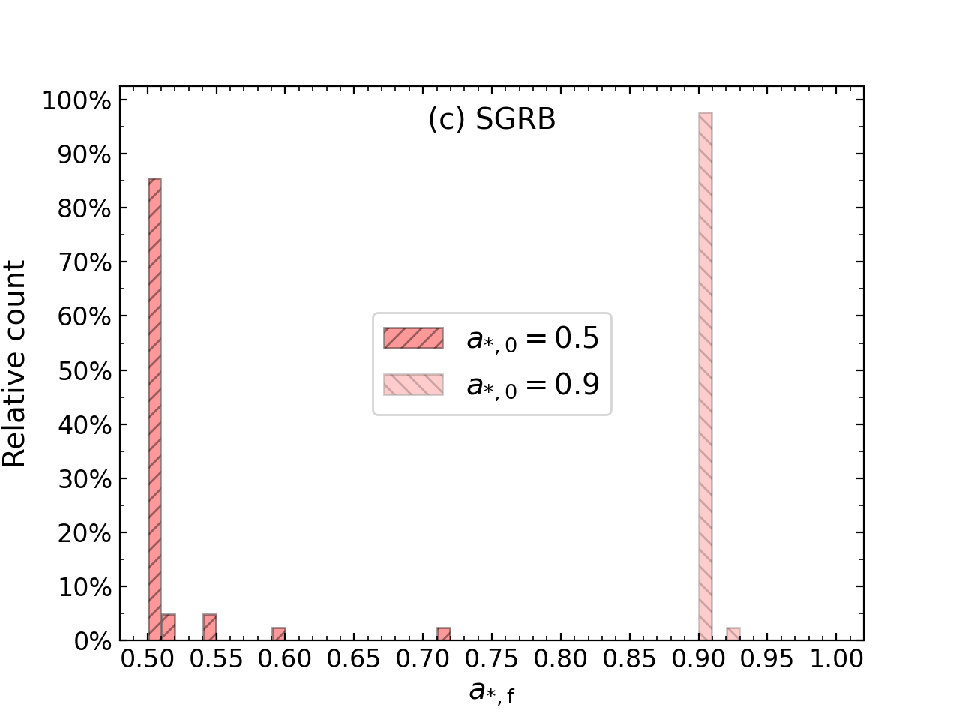}
\caption{Distributions of the final BH spin parameters $a_{\rm *,f}$. The dark and light colors denote $a_{*,0}= 0.5$ and $0.9$, respectively.}
\label{MyFigC}
\end{figure}

The mean luminosity of the GRB jet can be estimated as \citep[e.g.,][]{2011ApJ...739...47F,2015ApJ...806...58L}
\begin{equation}
L_{\mathrm{j}} \simeq \frac{\left(E_{\gamma, \mathrm{iso}}+E_{\mathrm{k}, \mathrm{iso}}\right)(1+z) \theta_{\mathrm{j}}^{2}}{2 T_{90}},
\label{eq3}
\end{equation}
where $E_{\gamma, \mathrm{iso}}$ is the isotropic radiated energy in the prompt emission phase, $E_{\mathrm{k}, \mathrm{iso}}$ is the isotropic kinetic energy powering long-lasting afterglow, $z$ is the redshift, $\theta_{\mathrm{j}}$ is the half-opening angle of the jet, and $T_{90}$ can be approximately considered as the duration of the violent activity of the central engine for both LGRBs and SGRBs. Using these observational GRB data, the jet luminosity $L_{\mathrm{j}}$ can be obtained. From the data of the prompt emission and afterglows, one can obtain the value of $E_{\gamma, \mathrm{iso}}$ and $E_{\mathrm{k}, \mathrm{iso}}$, and the radiative efficiency in the fireball model is implied in the fitting process. Generally, $E_{\mathrm{k}, \mathrm{iso}}$ is larger than $E_{\gamma, \mathrm{iso}}$ in most of cases as shown in Tables \ref{MyTabA}, \ref{MyTabB}, and \ref{MyTabC}. Moreover, in fireball model \citep[e.g.,][]{1992MNRAS.258P..41R,1993ApJ...405..278M}, the energy is released by the central engine in a very short timescale. The collisions of the shells in the internal shock phase will trigger the prompt emission and the external shocks sweeping the circumstellar media will power the long-lasting multiband afterglows. $T_{90}$ is considered to reflect the activity timescale of the GRB central engines.

For a BH hyperaccretion system in the centre of GRBs with the BZ mechanism, the energy output is determined by the BZ jet power $L_{\mathrm{BZ}}$. It is related to the BH mass accretion rate and the spin parameter $a_{*}$, which writes as \citep[e.g.,][]{2018ApJ...852...20L,2021ApJ...908..242D}
\begin{equation}
L_{\mathrm{BZ}}=9.3 \times 10^{53} a_{*}^{2} \dot{m} X\left(a_{*}\right)~\mathrm{erg} ~\mathrm{s}^{-1},
\label{eq4}
\end{equation}
where $X\left(a_{*}\right)=F\left(a_{*}\right) /(1+\sqrt{1-a_{*}^{2}})^{2}$, with $F\left(a_{*}\right)=\left[\left(1+q^{2}\right) / q^{2}\right][(q+1 / q) \arctan (q)-1]$ and $q=a_{*} /(1+\sqrt{1-a_{*}^{2}})$. Parameters $m_{\rm BH}=M_{\rm BH}/M_\odot$ and $\dot{m}=\dot{M}/M_\odot ~\rm s^{-1}$ are introduced in this work.

If we assume $L_{\rm j}=L_{\rm BZ}$ , the dimensionless mass accretion rate can be obtained as the function of $L_{\mathrm{j}}$:
\begin{equation}
    \dot{m}= \frac{L_{\mathrm{j}}}{9.3 \times 10^{53} a_{*}^{2} X\left(a_{*}\right)~\mathrm{erg} ~\mathrm{s}^{-1}}.
\label{eq5}
\end{equation}

According to the above equations, the final BH characteristics can be described. With a given initial BH mass (e.g. $M_{\rm BH,0}=3\,M_\odot$) and spin ($a_{*,0}=0.5$ and $0.9$) and the luminosity of a GRB jet $L_{\mathrm{j}}$ obtained by Equation~(\ref{eq3}), one can obtain the BZ jet power $L_{\mathrm{BZ}}$ and the values of $\dot{m}$ based on Equations~(\ref{eq4}) and (\ref{eq5}), respectively. Incorporating $\dot{m}$ and $L_{\mathrm{BZ}}$ into the Equations~(\ref{eq1}) and (\ref{eq2}), the mass and spin of the BH at the next time step can be solved. Step by step, the final BH mass $M_{\rm BH,f}$ and spin $a_{*,\rm f}$ can be obtained when the time reaches $T_{90}$. The fourth-order Runge-Kutta method, ode45 function of MATLAB, is used in our calculations. The main results are discussed below.

\section{Results} \label{sec:results}

In this work, we adopt the data of 206 LGRBs, 7 ULGRBs, and 41 SGRBs to calculate the evolution of BH mass and spin in the BH hyperaccretion system with the BZ mechanism. The duration $T_{90}$, redshifts $z$, half-opening angles $\theta_{\mathrm{j}}$, and isotropic radiated energy in the prompt emission phase $E_{\gamma, \mathrm{iso}}$ and isotropic kinetic energy $E_{\mathrm{k}, \mathrm{iso}}$ of the LGRBs, ULGRBs and SGRBs are collected in Tables~\ref{MyTabA}, \ref{MyTabB}, and \ref{MyTabC}, respectively. The GRBs associated with SNe are labelled by the superscript ``$*$''. The GRB luminosity derived from Equation (\ref{eq3}) is also presented. It should be noticed that the measurements of $E_{\rm \gamma,\rm iso}$ and  $E_{\rm k,\rm iso}$ are model dependent. Besides, the lower limits of $\theta_{\rm j}$ are used in some bursts.

Before the discussion of the GRB cases collected, theoretical results of mass growths with different initial BH masses and spin parameters are illustrated in Figure~\ref{MyFigA} based on our model. In the figure, we adopt a typical GRB luminosity $L_{\mathrm{j}}=10^{49} \,\mathrm{erg} \,\mathrm{s}^{-1}$ and duration $T_{90}=30\,\rm s$ to calculate the BH mass growths with different initial BH masses, i.e., $3,~5$, and $10\,M_\odot$ and different initial spin parameters, i.e., $a_{*,0} = 0.5$ and $0.9$. One can find that the final BH mass is almost independent of the initial BH mass ($\leq 10~M_\odot$) but depend on the initial spin and the BH growth will be more efficient with a smaller spin parameter. The conclusion can also be directly inferred from Equation~(\ref{eq5}). For a given luminosity, the initial BH mass has no effect on the accretion rate and then on the mass growth. However, the accretion rate is relevant to the BH spin, the larger initial spin parameter leads to a lower accretion rate, and thus a less final mass. As shown in Figure~\ref{MyFigA}, one can see that the mass growth is about $6 \times 10^{-3}\,M_{\odot}$ even with $a_{*,0}$ = 0.5. There is almost no change for the evolution of BH mass with the BZ mechanism for the typical luminosity and duration, and thus one can expect that the BH growths will be obvious only for the case with high GRB luminosity and long accretion timesacle.

The BH mass before accretion might be around $3\,M_{\odot}$ for the scenarios of the collapsars or the compact object mergers \citep[e.g.,][]{2017PhRvL.119p1101A,2021ApJ...915L...5A,2008ApJS..174..223B,2021ApJ...908..106L,2021MNRAS.507..431W,2022ApJ...929...83Q}. In the collapsar scenario, if the NDAF phase lasts several to tens of seconds, the central BH should be fed to grow up, but the disc outflows will decrease the BH's appetite, and more important, the jets might be chocked in the envelope and no observable electromagnetic signals. For merger scenarios, the jet may also be chocked by the tidal and/or post-merger ejecta \citep[e.g.,][]{2021ApJ...908..152M}. Considering the dissipative energy by the chocked jets to build cocoon, the total GRB energy should be less than the real released energy and the BH mass growth will be more violent. Here we just investigate the BH evolution after the jets break out from the progenitors. Combining with the result that the initial BH mass has almost no effect on the mass growths as shown in Figure 1, we use $M_{\rm BH,0}=3\,M_{\odot}$ in the GRBs sample calculations. Besides, $a_{*,0} = 0.5$ and $0.9$ are adopted.

\subsection{LGRB case}

The distributions of final BH masses $M_{\rm BH,f}$ and spins $a_{\rm *,f}$ for LGRB cases are presented in Figures~\ref{MyFigB}(a) and \ref{MyFigC}(a), respectively. The dark and light colors denote the initial BH spins $a_{*,0}= 0.5$  and $0.9$. As mentioned above, the mass growth under the BZ mechanism is inefficient. Thus, high GRB luminosity and long accretion timescale are required in this case. Moreover, a smaller initial BH spin parameter is favored.

Figure~\ref{MyFigB}(a) shows the final BH masses distribution after the accretion phase for 206 LGRB events. If the initial BH mass $M_{\rm BH,0}$ is 3 $M_\odot$, the mean final BH masses are about $3.203\,M_{\odot}$ for $a_{*,0}=0.5$ and  $3.059\,M_{\odot}$ for $a_{*,0}=0.9$. One can find that a smaller $a_{*,0}$ leads to a higher $M_{\rm BH,f}$ and that most of the final BH masses are below $ 5\,M_{\odot}$. With relatively high jet luminosity and long accretion time, one bursts, namely, GRB~050401, has a final BH mass larger than $5\,M_{\odot}$, which are $7.992\,M_{\odot}$ for $a_{*,0}=0.5$ and $7.172\,M_{\odot}$ for $a_{*,0}=0.9$, respectively, for which the BZ mechanism is a plausible prescription.

In Figure~\ref{MyFigC}(a), we show the distribution of the final BH spins for the LGRB events. The mean final BH spin parameters are about $0.60$ for $a_{*,0}=0.5$ and $0.91$ for $a_{*,0}=0.9$. The upper limit of the BH spin is set as $0.998$ \citep[e.g.,][]{2008bhad.book.....K}. The bursts with larger final BH masses have higher final BH spin parameters, and several bursts including the GRB~050401 mentioned above, can reach up to the spin limitation.

In our calculation GRB~050401 is a special case with high luminosity, which is mainly caused by the relatively high kinetic energy. Actually, some other bursts have the similar or even higher kinetic energy but with lower luminosity because of other parameters, such as the opening angle and $T_{90}$. Besides, it should be mentioned that the fitting of the kinetic energy is model-dependent \citep[e.g.,][]{2007ApJ...655..989Z}. We noticed that \citet[][]{2009MNRAS.394..214K} used a double-jet model to fit the X-ray afterglow of GRB~050401, the kinetic energies of the narrow and wide jets are $\sim 5.46 \times 10^{53}~\rm erg$ and $>5.07 \times 10^{51}~\rm erg$, respectively, which are much less than the value adopted here.

It is widely accepted that LGRBs signal the collapse of massive stars, which usually end their lives companied with SNe. The conclusion is confirmed by the observation evidence, such as the appearance of SN-like bumps in the optical afterglow light curves of several bursts \citep[e.g.,][]{2003Natur.423..847H,2009ApJ...703.1696Z}. There is competition on the luminosity of LGRBs and those of corresponding SN bumps \citep[e.g.,][]{2019ApJ...871..117S,2021ApJ...908..106L}. Thus, the typical luminosity of LGRBs associated with SNe is lower than that of LGRBs without observable SNe, which will further decrease the values of the BH growths. We indeed find that the BH growths are less obviously for LGRBs associated with SNe in most cases. Besides, it should be noticed that the SN explosions may exist in the cases without SNe observed in result of that they are too weak (even failed) or too distant to be observed.

\begin{center}
\begin{table*}
\caption{LGRB data}
\label{MyTabA}
\begin{tabular}{cccccccc}
\hline
\hline
GRB & $z$& $T_{90}$ &$\theta_{\rm j}$ & $E_{\rm \gamma,\rm iso}$& $E_{\rm k,\rm iso}$&$L_{\rm j}$ & Ref \\
~ &~ & (s) & (rad) & ($10^{52}$\,\rm erg) & ($10^{52}$\,\rm erg)  & ($10^{49} \rm erg\rm\,s^{-1}$) &  \\
\hline
        970508 & 0.8349 & 14.0 & 0.3775  & 0.61 & 0.99 & 14.94  & 1 \\
        970828 & 0.96 & 160 & 0.1239  & 29 & 37.154 & 6.22  & 1 \\
        971214 & 3.418 & 31.23 & >0.0967  & 21 & 8.48 & 19.50  & 1 \\
        980425* & 0.01 & 23.3 & 0.192  & 0.000093 & 0.00631 & 0.01  & 2 \\
        980613 & 1.0964 & 42.0 & >0.2194  & 0.59 & 1.22 & 2.17  & 1 \\
        980703 & 0.966 & 76.0 & 0.1957  & 7.2 & 2.41 & 4.76  & 1 \\
        990123 & 1.61 & 63.3 & 0.064  & 229 & 534 & 64.43  & 1 \\
        990510 & 1.619 & 67.58 & 0.0586  & 17 & 13.16 & 2.01  & 1 \\
        990705 & 0.84 & 32.0 & 0.093  & 18 & 0.34 & 4.56  & 1 \\
        991208 & 0.706 & 68 & 0.12  & 12 & 2.2 & 2.57  & 3 \\
        991216 & 1.02 & 15.17 & 0.0798  & 67 & 36.64 & 43.94  & 1 \\
        000210 & 0.846 & 9.0 & >0.1194  & 14.9 & 0.5 & 22.52  & 1 \\
        000926 & 2.0387 & 1.30 & 0.1075  & 27.1 & 9.97 & 500.67  & 1 \\
        010222 & 1.4769 & 74.0 & 0.0559  & 81 & 22.79 & 5.43  & 1 \\
        011121* & 0.362 & 47 & 0.157  & 7.8 & 2.7 & 3.75  & 2 \\
        011211 & 2.14 & 51.0 & 0.1114  & 5.4 & 71.32 & 29.31  & 1 \\
        020405 & 0.69 & 60 & 0.0822  & 14.79 & 945.48 & 91.38  & 4 \\
        020813 & 1.25 & 89 & 0.0541  & 66 & 204.174 & 10.00  & 1 \\
        021004 & 2.3304 & 77.1 & 0.2211  & 3.3 & 8.35 & 12.30  & 1 \\
        021211* & 1 & 2.8 & 0.024$ \sim $0.0768  & 1.12 & 4 & 1.09  & 2 \\
        030226 & 1.986 & 100 & 0.0417  & 7.94 & 2392.17 & 62.31  & 4 \\
        030323 & 3.37 & 26 & 0.0326  & 3.2 & 1093.91 & 97.99  & 4 \\
        030329* & 0.17 & 22.3 & 0.089  & 1.33 & 6.31 & 1.59  & 2 \\
        031203* & 0.1 & 40 & 0.157  & 0.0167 & 0.138 & 0.05  & 2 \\
        050126 & 1.29 & 30 & 0.365  & 0.8 & 39.8 & 206.44  & 1 \\
        050315 & 1.95 & 96 & 0.0759  & 5.7 & 512.403 & 45.86  & 1 \\
        050318 & 1.4436 & 32 & 0.038  & 2.2 & 11.259 & 0.74  & 1 \\
        050319 & 3.2425 & 139.4 & 0.038  & 4.6 & 77.896 & 1.81  & 1 \\
        050401 & 2.9 & 38 & 0.472  & 35 & 4570.9 & 52656.20  & 1 \\
        050416A & 0.654 & 5.4 & 0.237  & 0.1 & 15.1 & 130.75  & 1 \\
        050505 & 4.27 & 63 & 0.029  & 16 & 237.829 & 8.93  & 1 \\
        050525A* & 0.606 & 8.84 & 0.0551  & 2.5 & 8.2 & 2.95  & 2 \\
        050730 & 3.97 & 155 & >0.023  & 9 & 86.1223 & 0.81  & 1 \\
        050802 & 1.71 & 20 & 0.290  & 1.8197 & 616.6 & 3523.62  & 1 \\
        050803 & 0.42 & 87.9 & 0.333  & 0.25 & 37.5 & 33.81  & 5 \\
        050814 & 5.3 & 48 & 0.0419  & 6 & 831.764 & 96.52  & 1 \\
        050820A & 2.615 & 600 & 0.184 & 97.4 & 53.7145 & 15.41  & 1 \\
        050904 & 6.295 & 183.6 & 0.034  & 124 & 88.37 & 4.88  & 1 \\
        050922C & 2.198 & 4.5 & 0.026  & 5.3 & 47.725 & 12.74  & 1 \\
        051022 & 0.8 & 200 & 0.03  & 44 & 33.6 & 0.31  & 3 \\
        051109A & 2.35 & 360 & 0.0593  & 6.5 & 169.824 & 2.88  & 1 \\
        051109B & 0.08 & 14.3 & 0.4194  & 0.000679 & 0.02 & 0.14  & 5 \\
        051111 & 1.55 & 47 & 0.018  & 10.99 & 81.6 & 0.81  & 4 \\
        060115 & 3.53 & 139.6 & 0.126  & 6.05 & 85.68 & 23.63  & 5 \\
        060124 & 2.297 & 298 & 0.0531  & 41 & 578.87 & 9.67  & 1 \\
        060206 & 4.05 & 5 & 0.0351  & 4.3 & 386.76 & 243.30  & 1 \\
        060210 & 3.91 & 220 & 0.024  & 42 & 1313.2261 & 8.71  & 1 \\
        060418 & 1.49 & 52 & 0.029  & 13 & 7.5307 & 0.41  & 1 \\
        060502A & 1.51 & 28.4 & 0.3792  & 1.76 & 4.05 & 36.92  & 5 \\
        060505* & 0.089 & 4 & 0.4  & 0.0012 & 0.028 & 0.64  & 1 \\
        060526 & 3.21 & 258.8 & 0.063  & 2.6 & 15.58 & 0.59  & 1 \\
        060604 & 2.1357 & 95 & 0.1188  & 0.26 & 29.57 & 6.95  & 5 \\
        060605 & 3.8 & 19 & >0.046  & 2.5 & 115 & 31.41  & 1 \\
        060607A & 3.082 & 100 & >0.095  & 9 & 0.822 & 1.81  & 1 \\
        060614* & 0.12 & 6.9 & 0.2025  & 0.21 & 1.698 & 6.35  & 1 \\
        060707 & 3.42 & 210 & 0.1379  & 5.4 & 102.329 & 21.56  & 1 \\
        060708 & 1.92 & 10.2 & 0.332  & 0.96 & 0.52 & 23.35  & 5 \\
        060714 & 2.71 & 108.2 & 0.0201  & 7.7 & 250.46 & 1.79  & 1 \\
        060729 & 0.54 & 115.3 & 0.4687  & 0.53 & 15.56 & 23.61  & 5 \\
        060814 & 1.9229 & 145.3 & 0.1748  & 6.09 & 32.66 & 11.91  & 5 \\
        060904B & 0.703 & 192 & >0.174  & 0.72 & 9.4031 & 1.36  &6 \\
        060906 & 3.686 & 43.5 & 0.0684  & 9.13 & 28.23 & 9.41  & 5 \\
        060908 & 1.8836 & 18 & 0.008  & 9.8 & 2017.68 & 10.39  & 1 \\
        061007 & 1.262 & 75 & >0.138  & 86 & 29.9425 & 33.30  & 1 \\ \hline
\end{tabular}
\end{table*}

\begin{table*}
\contcaption{}
\label{tab:continued1}
\begin{tabular}{ccccccccc}
\hline

GRB & $z$& $T_{90}$ &$\theta_{\rm j}$ & $E_{\rm \gamma,\rm iso}$& $E_{\rm k,\rm iso}$&$L_{\rm j}$ & Ref \\

 & &  (s)  & (rad)& ($10^{52}$\rm erg) & ($10^{52}$\rm erg)  & ($10^{49} \rm erg\rm~s^{-1}$) &  \\
\hline
        061021 & 0.35 & 79 & 0.1501  & 10 & 6.166 & 3.11  & 1 \\
        061110A & 0.758 & 40.7 & 0.2532  & 0.44 & 0.38 & 1.14  & 5 \\
        061121 & 1.314 & 81 & 0.099  & 22.5 & 20.5215 & 6.02  & 1 \\
        061222A & 2.09 & 16 & 0.0471  & 21 & 2290.868 & 495.24  & 1 \\
        070110 & 2.352 & 89 & >0.274  & 3 & 0.687 & 5.21  & 1 \\
        070125 & 1.5477 & 63 & 0.2304  & 80.2 & 6.43 & 92.98  & 1 \\
        070129 & 2.3384 & 460.6 & 0.1422  & 8.06 & 42.92 & 3.74  & 5 \\
        070306 & 1.5 & 3 & 0.0768  & 6 & 67.608 & 180.90  & 1 \\
        070318 & 0.84 & 63 & 0.127  & 0.9 & 47.2719 & 11.35  & 1 \\
        070411 & 2.95 & 101 & 0.032  & 10 & 83.6596 & 1.88  & 1 \\
        070419A & 0.97 & 112 & >0.165  & 0.24 & 35.074 & 8.46  & 6 \\
        070508 & 0.82 & 23.4 & 0.0611  & 8 & 10.715 & 2.72  & 1 \\
        070529 & 2.4996 & 109.2 & 0.0996  & 4.7 & 17.06 & 3.46  & 5 \\
        071010A & 0.98 & 6 & 0.09  & 0.13 & 7.2164 & 9.82  & 1 \\
        071010B & 0.947 & 35.7 & 0.15  & 1.7 & 7.2713 & 5.50  & 1 \\
        071031 & 2.692 & 150.5 & 0.07  & 3.9 & 1.554 & 0.33  & 1 \\
        080109* & 0.007 & 600 & >0.1411  & 0.0000013 & 0.000446 & 0.0000075 & 7 \\
        080129 & 4.394 & 48 & >0.097  & 7 & 29.138 & 19.11  & 6 \\
        080310 & 2.43 & 32 & 0.0628  & 6.0256 & 29.512 & 7.51  & 1 \\
        080319C & 1.95 & 29.55 & >0.102  & 14.1 & 74.4078 & 45.96  & 1 \\
        080330 & 1.51 & 61 & >0.087  & 0.21 & 21.0923 & 3.32  & 1 \\
        080413B & 1.1 & 8 & 0.1047  & 2.4 & 138.038 & 202.06  & 1 \\
        080430 & 0.767 & 16.2 & 0.2842  & 0.45 & 8.37 & 38.85  & 5 \\
        080516 & 3.2 & 5.8 & 0.0364  & 0.98 & 35.5 & 17.50  & 5 \\
        080603A & 1.688 & 150 & 0.071  & 2.2 & 52.5129 & 2.47  & 1 \\
        080605 & 1.6398 & 20 & 0.0968  & 11.88 & 6.9 & 11.61  & 5 \\
        080707 & 1.23 & 27.1 & 0.1328  & 0.33 & 1.64 & 1.43  & 5 \\
        080710 & 0.845 & 120 & >0.062  & 0.8 & 2.6451 & 0.10  & 1 \\
        080721 & 2.591 & 16.2 & 0.1252  & 69.31 & 78.49 & 256.78  & 5 \\
        080810 & 3.35 & 108 & >0.105  & 45 & 41.8519 & 19.28  & 1 \\
        080905B & 2.374 & 128 & 0.1318  & 5.76 & 16.44 & 5.08  & 5 \\
        080913 & 6.695 & 8 & 0.359  & 8.6 & 10 & 1152.90  & 1 \\
        081007* & 0.53 & 9.01 & >0.349  & 0.15 & 0.15 & 3.10  & 2 \\
        081008 & 1.967 & 162.2 & 0.0227  & 9.98 & 134.7 & 0.68  & 1 \\
        081203A & 2.1 & 223 & >0.116  & 35 & 11.2261 & 4.32  & 1 \\
        081221 & 2.26 & 34 & 0.118  & 34.3 & 8.03 & 28.26  & 5 \\
        081222 & 2.77 & 5.8 & 0.0489  & 30 & 131.826 & 125.76  & 1 \\
        090313 & 3.375 & 78 & >0.093  & 3.2 & 276.8523 & 67.93  & 1 \\
        090323 & 3.568 & 133.1 & 0.0489  & 410 & 116 & 21.58  & 1 \\
        090328 & 0.7354 & 57 & 0.0733  & 13 & 82 & 7.77  & 1 \\
        090407 & 1.4485 & 310 & 0.0964  & 0.94 & 690 & 25.36  & 5 \\
        090418A & 1.608 & 56 & 0.0881  & 17.06 & 12.57 & 5.36  & 5 \\
        090423* & 8.23 & 10.3 & 0.0262  & 11 & 340 & 107.96  & 1 \\
        090424 & 0.544 & 49.47 & >0.378  & 4.6 & 53.1215 & 128.71  & 1 \\
        090516 & 4.109 & 210 & 0.0762  & 69.02 & 75.01 & 10.17  & 5 \\
        090529 & 2.625 & 100 & 0.0981  & 1.41 & 91.37 & 16.18  & 5 \\
        090530 & 1.266 & 48 & 0.1795  & 0.68 & 3.19 & 2.94  & 5 \\
        090618 & 0.54 & 113.2 & 0.2817  & 13.39 & 12.72 & 14.09  & 5 \\
        090812 & 2.452 & 75.9 & >0.071  & 40.3 & 148.827 & 21.68  & 1 \\
        090902B & 1.8829 & 19.328 & 0.0681  & 1.77 & 56 & 19.98  & 1 \\
        090926A & 2.1062 & 20 & 0.1571  & 210 & 6.8 & 415.51  & 1 \\
        091018 & 0.97 & 106.5 & 0.082  & 0.5888 & 12.023 & 0.78  & 1 \\
        091020 & 1.71 & 65 & 0.1204  & 12.2 & 51.286 & 19.18  & 1 \\
        091029 & 2.752 & 39.2 & >0.192  & 7.4 & 40.303 & 84.16  & 1 \\
        091127* & 0.48 & 68.7 & 0.096 & 4.3 & 22.9 & 2.70  & 2 \\
        091208B & 1.063 & 71 & 0.1274  & 2.01 & 50.119 & 12.29  & 1 \\
        100418A & 0.6235 & 8 & 0.356  & 0.99 & 3.36 & 55.94  & 1 \\
        100425A & 1.755 & 37 & 0.1565  & 0.49 & 1.32 & 1.65  & 5 \\
        100615A & 1.398 & 39 & 0.0791  & 5.82 & 16.66 & 4.32  & 5 \\
        100621A & 0.542 & 63.6 & >0.234  & 4.37 & 111.7596 & 77.09  & 1 \\
        100704A & 3.6 & 197.5 & 0.1159  & 17.75 & 29.45 & 7.38  & 5 \\
        100728B & 2.106 & 12.1 & >0.063  & 2.66 & 95.665 & 50.09  & 1 \\
        100814A & 1.44 & 174.5 & 0.1338  & 15.01 & 6160 & 772.88  & 5 \\
        100901A & 1.408 & 439 & 0.152  & 6.3 & 167.3233 & 11.00  & 1 \\
        100906A & 1.727 & 114.4 & 0.055  & 28.9 & 23.8173 & 1.90  & 1 \\
\hline
\end{tabular}
\end{table*}
\begin{table*}
\contcaption{}
\label{tab:continued2}
\begin{tabular}{ccccccccc}
\hline

GRB & $z$& $T_{90}$ &$\theta_{\rm j}$ & $E_{\rm \gamma,\rm iso}$& $E_{\rm k,\rm iso}$&$L_{\rm j}$ & Ref \\
 & &  (s)  & (rad)& ($10^{52}$\rm erg) & ($10^{52}$\rm erg)  & ($10^{49} \rm erg\rm~s^{-1}$) &  \\
\hline

        101219B* & 0.552 & 51 & >0.298  & 0.34 & 6.4 & 9.11  & 2 \\
        110205A & 2.22 & 257 & 0.064  & 56 & 31.2172 & 2.24  & 1 \\
        110213A & 1.46 & 48 & >0.142  & 6.9 & 25.7527 & 16.87  & 1 \\
        110715A & 0.82 & 13 & 0.2146  & 2.79 & 1.81 & 14.83  & 5 \\
        110808A & 1.35 & 48 & 0.1894  & 17.86 & 0.88 & 16.46  & 5 \\
        111008A & 4.9898 & 63.46 & 0.1059  & 42.11 & 23.77 & 34.87  & 5 \\
        111123A & 3.1516 & 290 & 0.0544  & 23.78 & 89.24 & 2.39  & 5 \\
        111215A & 2.06 & 374 & 0.03  & 14 & 770 & 2.89  & 3 \\
        111228A & 0.714 & 101.2 & 0.259  & 4.17 & 13.89 & 10.26  & 5 \\
        120119A & 1.728 & 70 & 0.032  & 36 & 4.17 & 0.80  & 1 \\
        120326A & 1.798 & 11.8 & 0.0803  & 3.2 & 14 & 13.15  & 1 \\
        120327A & 2.813 & 62.9 & 0.0694  & 8.78 & 8.45 & 2.52  & 5 \\
        120404A & 2.876 & 48 & >0.024  & 9 & 7.8 & 0.39  & 6 \\
        120422A* & 0.283 & 5.35 & 0.2  & 0.0045 & 0.12 & 0.60  & 2 \\
        120521C & 6 & 26.7 & 0.0524  & 8.25 & 22 & 10.89  & 1 \\
        120712A & 4.1745 & 14.7 & 0.0467  & 18.57 & 8.69 & 10.46  & 5 \\
        120729A & 0.8 & 72 & 0.0204  & 1.24 & 716.77 & 3.74  & 4 \\
        120811C & 2.671 & 26.8 & 0.0486  & 6.96 & 27.8 & 5.62  & 5 \\
        120815A & 2.358 & 9.7 & >0.063  & 2.12 & 7.08 & 6.32  & 6 \\
        120922A & 3.1 & 173 & 0.0977  & 21.38 & 59.62 & 9.16  & 5 \\
        120923A & 8.1 & 26.1 & 0.0855  & 4.8 & 0.29 & 6.49  & 8 \\
        121024A & 2.298 & 69 & 0.0741  & 1.84 & 32.61 & 4.52  & 5 \\
        121128A & 2.2 & 23.3 & 0.0524  & 1.05 & 14.73 & 2.98  & 5 \\
        121211A & 1.023 & 182 & 0.1026  & 0.17 & 37.76 & 2.22  & 5 \\
        130215A* & 0.597 & 66 & >0.1772  & 3.1 & 12.3 & 5.85  & 7 \\
        130420A & 1.297 & 123.5 & 0.2103  & 0.23 & 3.72 & 1.62  & 5 \\
        130427A* & 0.34 & 163 & >0.0873  & 96 & 13.1 & 3.42  & 2 \\
        130606A & 5.913 & 276.58 & 0.0498  & 21.67 & 86.86 & 3.36  & 5 \\
        130612A & 2.006 & 4 & 0.0547  & 0.76 & 3.19 & 4.44  & 5 \\
        130702A* & 0.145 & 59 & 0.086  & 0.064 & 37.7 & 2.71  & 2 \\
        130831A* & 0.479 & 32.5 & $\geqslant$0.123  & 0.46 & 11.4 & 4.08  & 2 \\
        130907A & 1.238 & 364 & 0.01  & 8 & 441 & 0.14  & 3 \\
        131030A & 1.295 & 41.1 & 0.2149  & 17.22 & 6.73 & 30.88  & 5 \\
        131105A & 1.686 & 112.3 & 0.1284  & 34.31 & 22.24 & 11.15  & 5 \\
        140206A & 2.73 & 93.6 & 0.1291  & 30.68 & 46.28 & 25.56  & 5 \\
        140311A & 4.954 & 70 & 0.03  & 10 & 2270 & 87.27  & 3 \\
        140423A & 3.26 & 134 & >0.3  & 65.4 & 2140 & 3155.04  & 9 \\
        140512A & 0.725 & 154.8 & 0.093  & 9.07 & 51.44 & 2.92  & 5 \\
        140518A & 4.707 & 60.5 & 0.0277  & 4.99 & 8.41 & 0.48  & 5 \\
        140606B* & 0.384 & 23.6 & 0.14  & 0.347 & 0.039 & 0.22  & 2 \\
        140629A & 2.276 & 42 & 0.04 & 4.4 & 1800 & 112.59  & 10 \\
        140703A & 3.14 & 67.1 & 0.0285  & 1.95 & 1260 & 31.62  & 5 \\
        140903A & 0.351 & 0.3 & 0.03  & 0.006 & 47.3 & 95.87  & 3 \\
        141121A & 1.47 & 549.9 & 0.0635  & 0.02 & 54200 & 490.83  & 5 \\
        150403A & 2.06 & 40.9 & 0.1101  & 1.51 & 248.8 & 113.51  & 5 \\
        150910A & 1.359 & 112.2 & 0.0546  & 2.06 & 875 & 27.49  & 5 \\
        151027A & 0.81 & 129.69 & 0.1562  & 3.38 & 28.71 & 5.46  & 5 \\
        151215A & 2.59 & 17.8 & 0.0972  & 0.55 & 1.16 & 1.63  & 5 \\
        160121A & 1.96 & 12 & 0.0331  & 0.77 & 208 & 28.21  & 5 \\
        160131A & 0.97 & 325 & 0.0524  & 83 & 50 & 1.11  & 11 \\
        160227A & 2.38 & 316.5 & 0.1711  & 4.06 & 12.76 & 2.63  & 5 \\
        160327A & 4.99 & 28 & 0.0357  & 8.9 & 10.38 & 2.63  & 5 \\
        160509A & 1.17 & 371 & 0.04  & 86 & 76.5 & 0.76  & 3 \\
        160625B & 1.406 & 460 & 0.041  & 300 & 194 & 2.17  & 3 \\
        161108A & 1.159 & 105.1 & 0.206  & 0.54 & 0.66 & 0.52  & 5 \\
        161117A & 1.549 & 125.7 & 0.1525  & 23.29 & 42.15 & 15.43  & 5 \\
        161219B* & 0.1475 & 10 & 0.6981  & 0.0097 & 0.016 & 0.72  & 7 \\
        170113A & 1.968 & 20.66 & 0.1501  & 0.63 & 8.07 & 14.08  & 5 \\
        171010A & 0.33 & 104 & 0.06  & 22 & 10.5 & 0.75  & 3 \\
        181110A & 1.505 & 138.4 & 0.065 & 13.43 & 166 & 6.86  & 12 \\
        181201A & 0.45 & 19.2 & $\gtrsim$0.1047  & 12 & 22 & 14.07  & 13 \\
        190106A & 1.861 & 76.8 & 0.0733  & 10 & 9 & 1.90  & 14 \\
        190114C & 0.425 & 116.354 & >0.5672  & 41.2 & 190 & 455.47  & 15 \\
        190530A & 0.9386 & 20.31 & 0.0665  & 143 & 3715.4 & 814.33  & 16 \\
        190829A & 0.0785 & 57 & 0.015 & 27 & 40 & 0.14  & 17 \\
        \hline
        \end{tabular}
\end{table*}

\begin{table*}
\contcaption{}
\label{tab:continued3}
\begin{tabular}{ccccccccc}
\hline
GRB & $z$& $T_{90}$ &$\theta_{\rm j}$ & $E_{\rm \gamma,\rm iso}$& $E_{\rm k,\rm iso}$&$L_{\rm j}$ & Ref \\
 & &  (s)  & (rad)& ($10^{52}$\rm erg) & ($10^{52}$\rm erg)  & ($10^{49} \rm erg\rm~s^{-1}$) &  \\
\hline

        191221B & 1.148 & 13 & 0.0454  & 36 & 9.4 & 7.73  & 18 \\
        200826A* & 0.7481 & 0.65 & 0.24 & 0.717 & 6 & 520.26  & 19 \\
        200829A & 1.25 & 6.9 & 0.09 & 141 & 44.668 & 245.20  & 20 \\
        201015A & 0.423 & 10 & 0.9772  & 0.011 & 1.6218 & 110.94  & 21 \\
        201216C & 1.1 & 48 & 0.123  & 57.6 & 1122 & 390.38  & 21 \\
        210104A & 0.46 & 35 & 1.0  & 1.3 & 0.39 & 35.25  & 22 \\
        210204A* & 0.876 & 207 & 0.024  & 114.8 & 114.82 & 0.60  & 23 \\
        210205A & 2.514 & 22.7 & 0.96 & 1.52 & 2.3988 & 279.54  & 24 \\
        210731A & 1.2525 & 22.5 & $\gtrsim$0.1047  & 1.29 & 63 & 35.28  & 25 \\
        210905A & 6.312 & 870 & 0.147 & 127 & 508 & 57.66  & 26 \\
        220101A & 4.618 & 128 & 0.0229  & 300 & 34674 & 402.49  & 27 \\
        221009A & 0.151 & 327 & 0.024 & 1500 & 676.08 & 2.21  & 28 \\ \hline\hline
\end{tabular}
\begin{minipage}{16cm}
\emph{Notes}: \\
$^{\star}$ GRBs with SNe observed.\\
\emph{References}: \\
(1) \citet{2018MNRAS.477.2173S};
(2) \citet{2019ApJ...871..117S};
(3) \citet{2021ApJ...911...14K};
(4) \citet{2018ApJ...859..160W};
(5) \citet{2018ApJS..236...26L};
(6) \citet{2018ApJ...862..130L} ;
(7) \citet{2017JHEAp..13....1Y};
(8) \citet{2018ApJ...865..107T};
(9) \citet{2020ApJ...900..176L};
(10) \citet{2018ApJ...860....8X};
(11) \citet{2022A&A...658A..11M};
(12) \citet{2022Univ....8..248H};
(13) \citet{2019ApJ...884..121L};
(14) \citet{2023ApJ...948...30Z} ;
(15) \citet{2021ApJ...908L...2S};
(16) \citet{2022RAA....22f5002L};
(17) \citet{2021MNRAS.504.5647S};
(18) \citet{2023NatAs...7...80U};
(19) \citet{2021NatAs...5..917A};
(20) \citet{2023ApJ...944...21L};
(21) \citet{2023ApJ...952..127Z};
(22) \citet{2022ApJ...941...63Z};
(23) \citet{2022MNRAS.513.2777K};
(24) \citet{2022JApA...43...11G};
(25) \citet{2023A&A...671A.116D};
(26) \citet{2022A&A...665A.125R};
(27) \citet{2022ApJ...941...82M};
(28) \citet{2023ApJ...947...53R}.
\end{minipage}
\end{table*}
\end{center}

\begin{table*}
\caption{ULGRB Data}
\centering
\begin{tabular}{cccccccc}
\hline
\hline
GRB & $z$& $T_{90}$ &$\theta_{\rm j}$ & $E_{\rm \gamma,\rm iso}$& $E_{\rm k,\rm iso}$&$L_{\rm j}$ & Ref \\
~ &~ & (s) & (rad) & ($10^{52}$\rm erg) & ($10^{52}$\rm erg)  & ($10^{49} \rm erg\rm~s^{-1}$) &  \\ \hline

    	060218A*  & 0.0331 & 2100 & $\gtrsim1.4$  & 0.0062 & 0.0001 & 0.0030   & 1 \\
	091024 & 1.092 & 1020 & >0.071  & 28 & 37.2529 & 0.34  & 2 \\
        100316D*  & 0.0591 & 1300 & >1.3963  & 0.006 & 0.000605 & 0.0052   & 3 \\
         101225A  & 0.847 & 1090 & 0.2094  & 10 & 24 & 1.2631   & 4 \\
        111209A*  & 0.677 & 15000 & >0.2094  & 96 & 59 & 0.3799   & 4 \\
        121027A  & 1.773 & 1070 & 0.2094  & 140 & 15 & 8.8069   & 4 \\
        130925A & 0.35 & 4500 & 0.0012  & 15 & 22.779 & 8.46E-06 & 5,6 \\ \hline
\end{tabular}
\begin{minipage}{16cm}
\emph{References}: \\
(1) \citet{2006Natur.442.1014S};
(2) \citet{2018MNRAS.477.2173S};
(3) \citet{2013ApJ...778...18M};
(4) \citet{2013ApJ...778...67N};
(5) \citet{2014MNRAS.444..250E};
(6) \citet{ 2015ApJ...812...86H}.\\
\end{minipage}
\label{MyTabB}
\end{table*}

\begin{table*}
\centering
\caption{SGRB Data}
\begin{tabular}{cccccccc}
\hline
\hline
GRB & $z$& $T_{90}$ &$\theta_{\rm j}$ & $E_{\rm \gamma,\rm iso}$& $E_{\rm k,\rm iso}$&$L_{\rm j}$ & Ref \\
~ &~ & (s) & (rad) & ($10^{52}$\rm erg) & ($10^{52}$\rm erg)  & ($10^{49} \rm erg\rm~s^{-1}$) &  \\\hline
	050509B & 0.225 & 0.04 & >0.05 & 0.00024 & 0.0055 & 0.22  & 1 \\
        050709 & 0.161 & 0.07 & >0.26 & 0.0027 & 0.0016 & 2.41  & 1 \\
        050724A & 0.257 & 3 & >0.35 & 0.009 & 0.027 & 0.92  & 1 \\
        051210 & 1.3 & 1.3 & >0.05 & 0.4 & 0.238 & 1.41  & 1 \\
        051221A & 0.5465 & 1.4 & 0.12 & 0.28 & 1.26 & 12.25  & 1 \\
        060502B & 0.287 & 0.09 & >0.05 & 0.003 & 0.012 & 0.27  & 1 \\
        060801 & 1.13 & 0.5 & 0.056 & 0.7 & 0.071 & 5.17  & 1 \\
        061006 & 0.438 & 0.4 & 0.41 & 3 & 0.314 & 986.76  & 1 \\
        061201 & 0.111 & 0.8 & 0.017 & 3 & 0.007 & 0.60  & 1 \\
        061210 & 0.409 & 0.2 & >0.37 & 0.09 & 0.086 & 84.87  & 1 \\
        070429B & 0.902 & 0.5 & >0.05 & 0.07 & 0.451 & 2.48  & 1 \\
        070714B & 0.923 & 2.0 & 0.33 & 1.16 & 0.232 & 72.88  & 1 \\
        070724A & 0.457 & 0.4 & 0.27 & 0.003 & 0.099 & 13.54  & 1 \\
        070729 & 0.8 & 0.9 & >0.05 & 0.017 & 0.132 & 0.37  & 1 \\
        070809 & 0.473 & 1.3 & 0.4 & 0.0056 & 0.391 & 35.95  & 1 \\
        071227 & 0.381 & 1.8 & >0.026 & 0.1 & 0.025 & 0.03  & 1 \\
        080905A & 0.122 & 1.0 & 0.28 & 0.0005 & 0.0024 & 0.13  & 1 \\
        090426 & 2.609 & 1.2 & 0.07 & 0.5 & 13.5 & 103.16  & 1 \\
        090510 & 0.903 & 0.3 & 0.017 & 4.47 & 0.307 & 4.38  & 1 \\
        090515 & 0.403 & 0.04 & >0.05 & 0.0008 & 0.062 & 2.75  & 1 \\
        100117A & 0.92 & 0.3 & 0.27 & 0.09 & 0.11 & 46.66  & 1 \\
        100206A & 0.408 & 0.1 & >0.05 & 0.0763 & 0.0073 & 1.47  & 1 \\
        100625A & 0.453 & 0.3 & >0.05 & 0.075 & 0.0093 & 0.51  & 1 \\
        101219A & 0.718 & 0.6 & 0.29 & 0.49 & 0.045 & 64.42  & 1 \\
        110731A & 2.83 & 7.3 & >0.15 & 68 & 7.5 & 472.06  & 2 \\
        111020A & 1.5 & 0.4 & 0.05$\sim$0.14 & 0.19 & 1.2 & 11.91  & 3 \\
        111117A & 1.3 & 0.5 & 0.11 & 0.338 & 0.377 & 18.13  & 1 \\
        120804A & 1.3 & 0.81 & >0.19 & 0.7 & 0.7 & 71.75  & 1 \\
        130603B & 0.356 & 0.18 & 0.07 & 0.212 & 0.28 & 9.08  & 1 \\
        131001A & 0.717 & 1.54 & >0.05 & 0.037 & 0.541 & 0.81  & 1 \\
        140622A & 0.959 & 0.13 & >0.05 & 0.0065 & 0.977 & 18.53  & 1 \\
        140903A & 0.351 & 0.3 & 0.09 & 0.006 & 4.3 & 78.53  & 4 \\
        150101B & 0.1343 & 0.012 &  $\gtrsim0.16$ & 0.0013 & 0.61 & 712.15  & 5\\
        160624A & 0.483 & 0.2 & 0.16 & 0.047 & 0.05  & 9.11  & 6 \\
        160821B & 0.16 & 0.48 & 0.063 & 0.021 & 8 & 38.47  & 1 \\
        170817A & 0.009783 & 0.5 & 0.2 & 0.0022 & 0.02 & 0.90  & 7 \\
        180418A & 0.5 & 1.504 & >0.122 & 0.3& 0.077 & 2.80 & 8 \\
        181123B & 1.754 & 0.26 & 1.57 & 0.026 & 0.234 & 3394.17  & 9 \\
        200522A & 0.5536 & 0.62 & 0.25 & 0.0084 & 0.015849 & 1.97  & 10 \\
        200826A* & 0.7481 & 1.14 & 0.24 & 0.717 & 6 & 296.64  & 11 \\
        211211A & 0.076 & 6 & 0.033 & 0.76 & 1.2882 & 0.20  & 12 \\ \hline
\end{tabular}
\begin{minipage}{16cm}
\emph{Notes}: \\
GRB 200826A is a short LGRB, which is associated with SNe.\\
\emph{References}: \\
(1) \citet{2018MNRAS.477.2173S};
(2) \citet{2017ApJ...843..114L};
(3) \citet{2012ApJ...756..189F};
(4) \citet{2016ApJ...827..102T};
(5) \citet{2016ApJ...833..151F};
(6) \citet{2021MNRAS.502.1279O};
(7) \citet{2018ApJ...854...60M};
(8) \citet{2019ApJ...881...12B};
(9) \citet{2021ApJ...911L..28D} ;
(10) \citet{2021ApJ...906..127F};
(11) \citet{2021NatAs...5..917A};
(12) \citet{2023ApJ...947L..21Z}.
\end{minipage}
\label{MyTabC}
\end{table*}

\subsection{ULGRB case}

ULGRBs are characterised by gamma-ray emission lasting longer than $\sim 1000 \mathrm{~s}$. Although their duration are different as typical LGRBs, the evidence for a new GRB classification is still unclear \citep[e.g.,][]{2014ApJ...787...66Z,2018ApJ...859...48P}. Even so, given by the long accretion time, we expect  correspondingly large BH growths and we thus separate them in Table~\ref{MyTabB}.

The distribution of the final BH masses after the accretion phase for ULGRB events is shown in Figure~\ref{MyFigB}(b). The figure shows that the mass growths are marginal. None of the bursts in the sample has a final BH mass larger than $5\,M_{\odot}$ even with $a_{*,0}=0.5$ for the initial BH mass $M_{\rm BH,0} =3 M_\odot$. In this case, the mean final BH masses are about $3.193\,M_{\odot}$ and  $3.039\,M_{\odot}$ for $a_{*,0}=0.5$ and $0.9$, respectively.

Figure~\ref{MyFigC}(b) presents the distribution of the final BH spins for ULGRB events. The evolution of spin parameters is also inefficient. The mean final BH spin parameters are about $0.61$ for $a_{*,0}=0.5$ and  $0.92$ for $a_{*,0}=0.9$.

One can find that the BH growths are inefficient for ULGRBs. This is because the long accretion time cannot offset the effect caused by the low burst luminosity.

\subsection{SGRB case}

Similar to LGRBs, the evolution of BH characteristic parameters is investigated for SGRBs. For the merger scenario, the ejecta hardly stops the SGRB jets, but the limited accretion matter can only support the central engine activity for no more than seconds. Thus, one can expect that the growths of BHs in the centre of SGRBs are undoubtedly very limited.

The distributions of the final BH masses and spin parameters of 41 SGRB events are presented in Figures~\ref{MyFigB}(c) and \ref{MyFigC}(c), respectively. We find that if the initial BH mass $M_{\rm BH,0} =3 M_\odot$, the mean final BH masses are about $3.0173\,M_{\odot}$ for $a_{*,0}=0.5$ and $3.0026\,M_{\odot}$ for $a_{*,0}=0.9$. Besides, the mean final BH spin parameters are about $0.512$ for $a_{*,0}=0.5$ and  $0.901$ for $a_{*,0}=0.9$. It shows that the effect on the evolution of BH characteristic parameters is not significant and the final BH masses for all of the bursts are between $3 \mbox{-}5\, M_{\odot}$.

As the results, the BH growths in SGRBs is consistent with the mass supply limitation in the scenario of compact object mergers.

\section{Conclusions and discussion} \label{sec:conclusions}

Based on the observational data, we test the capacity of the BZ mechanism to power GRBs and present the growths of BHs in the center of GRBs. The BZ mechanism is capable of powering all types of GRBs with reasonable parameters of BH hyperaccretion systems and the main conclusions are summarised as follows:

\begin{enumerate}
\item
By assuming GRBs powered by the BZ jets, for LGRBs with typical durations, the mean BH mass growth is about $0.203\,M_{\odot}$ for $a_{*,0}=0.5$. Only GRB~050401 has a final BH mass larger than $5\,M_{\odot}$ for the initial BH mass $\sim 3~M_\odot$. At the same time, the BH spin parameter for GRB~050401 almost reaches up to $0.998$.
\item
For ULGRBs with duration $T_{90} > 1000 \mathrm{~s} $, the mean BH mass growth is about $0.193 M_{\odot}$ for $a_{*,0}=0.5$.
\item
For SGRBs, the mean BH mass growths are about $0.0173 M_{\odot}$ for $a_{*,0}=0.5$ and $0.0026\,M_{\odot}$ for $a_{*,0}=0.9$.
\end{enumerate}

It should be noticed that the jet half-opening angle $\theta_{\mathrm{j}}$ is usually estimated by the jet break time in X-ray afterglows and it is hard to constrain since the absence of jet break observations. The lower limits of $\theta_{\mathrm{j}}$ are widely used \citep[e.g.,][]{2015ApJ...806...58L,2017JHEAp..13....1Y}, and therefore one can see from Equation (\ref{eq3}) that the BH growths are lower limits in this work.

The assumption of a Keplerian disc is adopted in this work. Actually, \citet[][]{2004ApJ...602..312G} investigated the BH spin evolution for the thick torus accretion, and they found that the spin-down effects will be resulted in case of large initial spins for some physical processes. Besides, \citet[][]{2008ApJ...687..433J} studied the evolution of the BH spin in LGRB cases and found that the spin-down of the BH is possible due to the low specific angular momentum of the accretion materials.

The Swift satellite observations have revealed the diverse morphology of light curves in X-ray afterglows \citep[e.g.,][]{2004ApJ...611.1005G,2005SSRv..120..165B}. Among them, X-ray flares are important signatures, which may be caused by the restart of central engine \citep[e.g.,][]{2005Sci...309.1833B,2017NewAR..79....1L}. The BZ mechanism is a promising candidate to explain the GRBs accompanied by long-lasting X-ray flares \citep[e.g.,][]{2013ApJ...773..142L}. If the effect of flares is considered, the evolution of BH will be more violent. In addition, the long-lasting plateau phases are observed in some GRB afterglows, and energy injection is one of the most widely accepted explanations \citep[e.g.,][]{2006MNRAS.372L..19F,2006ApJ...642..354Z}. In this case, the external shock will be refreshed by the huge energy contribution from the central engine. As well as X-ray flares, if the plateau phase induced by the energy injection is originated from the long-lasting BH hyperaccretion \citep[e.g.,][]{2021ApJ...916...71H}, the BH will undergo more drastic evolution to promote its growths.

Moreover, in collapsar and merger scenarios, the outflows from the BH hyperaccretion disc are definitely strong \citep[e.g.,][]{2018ApJ...852...20L} and will inject matter and energy into the shock to power luminous SNe and kilonovae \citep[e.g.,][]{2006ApJ...643.1057S,2018MNRAS.477.2173S,2019ApJ...871..117S,2019LRR....23....1M,2022ApJ...925...43Q}. Once the disc outflows are strong enough, the accretion rate in the inner region of the disc is too low to ignite NDAFs, which should weaken the domination of the neutrino annihilation process to power GRBs.

\section*{Acknowledgements}

We thank Hui-Min Qu, Shuang-Xi Yi, Shu-Jin Hou, Shu-Yu Hu, and Cui-Ying Song for the helpful discussion. This work was supported by the National Natural Science Foundation of China under grants 12173031 and 12221003.

\section*{Data availability}

The data underlying this article will be shared on reasonable request to the first author.

\end{document}